\documentclass{aastex62}

\def\submitted{ApJ, submitted.}
\submitjournal{AJ}

\shorttitle{SMC star clusters}
\shortauthors{Andr\'es E. Piatti}

\begin{document}

\title{Hints for multiple populations in intermediate-age clusters of the Small Magellanic Cloud }

\correspondingauthor{Andr\'es E. Piatti}
\email{andres@oac.unc.edu.ar}

\author{Andr\'es E. Piatti}
\affiliation{Consejo Nacional de Investigaciones Cient\'{\i}ficas y T\'ecnicas, Av. Rivadavia 1917, 
C1033AAJ,Buenos Aires, Argentina}
\affiliation{Observatorio Astron\'omico de C\'ordoba, Laprida 854, 5000,
C\'ordoba, Argentina}

\begin{abstract}
We report on the magnitude of the intrinsic [Fe/H] spread
in the Small Magellanic Cloud (SMC) intermediate-age massive clusters NGC\,339, 361, Lindsay\,1
and 113, respectively. In order to measure the cluster metallicity dispersions, we 
used accurate Str\"omgren 
photometry of carefully selected cluster red giant 
branch (RGB) stars. We  determined the Fe-abundance spreads by employing a maximum 
likelihood approach. The spreads obtained using the more accurate
photometry of the brighter RGB stars resulted to be marginal ($\sim$ 0.05$\pm$0.03 dex)
for NGC\,339 and NGC\,361, while for Lindsay\,1 and Lindsay\,113 we obtained 
metallicity spreads of 0.00$\pm$0.04 dex. 
From these results, we speculated with the possibility that NGC\,361 is added to the 
group of four SMC clusters with observational evidence of multiple populations (MPs). 
Furthermore, in the context of the present debate about the
existence of Fe-abundance inhomogeneities among old clusters with MPs, these outcomes 
put new constrains to recent theoretical speculations for making this phenomenon visible.
\end{abstract}


\section{Introduction}  

Four intermediate-age (5-8 Gyr) massive ($\ga 10^5 M_\odot$) 
clusters of the Small Magellanic Cloud (SMC)
have been found to harbour multiple stellar populations 
\citep[Kron\,3, Lindsay\,1, NGC\,339 and 416][]{hollyheadetal2018}; to which
other fourteen may also be added to the list of suitable targets for
multiple population (MP) searches, namely: Lindsay\,38 \citep{petal01}, Lindsay\,110, 112, 
113 \citep{petal07}, NGC\,361 \citep{metal98}, AM\,3, HW\,40, 41, 42, 59, 63 \citep{p11c}, 
HW\,22 \citep{p11b}, ESO\,51-SC9 \citep{p12b} and BS\,196 \citep{betal08b}.
A common feature that arises from Kron\,3 \citep{hollyheadetal2018} and Lindsay\,1
\citep{hollyheadetal2017} spectroscopic analyses is that the clusters show spreads in the abundances
of light elements, a characteristic commonly seen in old globular clusters.
For NGC\,339 and NGC\,416 \citet{niederhoferetal2017} obtained $HST$ $UV$/optical data from which
difference in the abundance of N can be inferred. As far as we are aware, there is
no study to seek for variation in Fe-abundances ([Fe/H]) in these SMC clusters.

The existence of [Fe/H] intrinsic spread among Milky Way globular clusters with
observed MPs seems to be a feature of a very small minority. Only 8 
Milky Way old globular clusters out of a population of 156 \citep{harris1996} have inhomogeneities
in [Fe/H] $>$ 0.05 dex \citep{johnsonetal2015,marinoetal2015}. 
Among the 15 known Large Magellanic Cloud old globular clusters, only NGC\,1754, 2210, and 2257 
have been searched for anomalies in their metallicities by \citet{mucciarellietal2009},
who concluded from [Fe/H] values of 5--7 stars per globular cluster that they exhibit quite homogeneous iron
 abundances
(intrinsic spread=0.02--0.04 dex) despite the observed occurrence of light element variations
 such as an Na-O anti-correlation. Therefore, 
they show very similar properties to the vast majority of Galactic globular clusters.
 From this perspective, anomalies  in the iron content would not appear 
to be the most frequent manifestation of the globular cluster MPs
formation, as recently reviewed by \citet{bl2017}.

From a theoretical point of view, different models have recently proposed distinct scenarios
to describe abundance anomalies in a variety of chemical elements in massive
clusters harboring MPs. For instance, \citet{bt2016}'s model is based on merger events,
\citet{bekki2017} and \citet{kl2018} used supernovae enrichment and asymptotic
giant branch star ejecta, while \citet{gielesetal2018} proposed a concurrent formation of
globular clusters and supermassive stars, among others.

These models have been mainly stimulated by observational findings of (anti-)correlations 
between chemical abundances of certain light elements 
\citep[e.g., Na-O, Mg-Al, Mg-Si, Si-Zn;][]{o1971,c1978,carrettaetal2009,grattonetal2012,hankeetal2017}
and  bimodalities in CN and CH that trace light element variations 
\citep[e.g.][and references therein]{kayseretal2008,mg2010}. 
Some of the models also propose mechanisms to obtain
intrinsic [Fe/H] spreads $>$ 0.05 dex. For instance, \citet{gavagninetal2016}'s model used
merger events, while that of \citet{bailin2018} is based on feedback from core-collapse supernovae.
Recently, \citet{limetal2017} found that globular clusters with large intrinsic [Fe/H] spreads
also show a positive CN-CH correlation.

As Fe-abundance spread is considered, a frequent idea relies on the
formation by merger events  of two massive clusters with initially different
[Fe/H], although there are still puzzling observational results 
which need to be addressed. \citet{gavagninetal2016} showed that the key parameters are
the initial mass and density ratios of the progenitors. For instance, they found that the 
radial distributions of MPs in $\omega$ Cen and NGC\,1851 are matched if the less
massive progenitor are four times as dense as the larger one.
The cluster's mass has also been assumed to be a
driver for such a metallicity inhomegeneity, in the sense that massive clusters
($>$10$^7M_\odot$) could inhibit the formation of second generation of stars with [Fe/H] 
by more than 0.05 dex larger than that for the first generation, because of SN 
feedback effects in the molecular clouds out of which globular clusters are formed
\citep{bekki2017}. 
Indeed, the least massive Milky Way globular cluster ESO\,452-SC11
\citep[$M=$6.8$\pm$3.4 $\times$10$^3M_\odot$][]{simpsonetal2017} does
show MPs of stars each with a different chemical composition.

In this work we made use of Str\"omgren photometry of the SMC intermediate-age massive
\citep[$\ga$10$^5M_\odot$][]{mvdm2005,getal11} clusters
NGC\,339, 361, Lindsay\,1 and 113 with the aim of investigating their metallicity
distributions. The Str\"omgren photometric system has resulted to be very efficient at 
identifying MPs in globular clusters \citep{massarietal2016,gruytersetal2017}. Its 
medium-bandwidth filters has the advantage of allowing us to obtain accurate [Fe/H] values
for many stars in a cluster field, straightforwardly, provided the photometric data
are precise. Particularly, \citet{franketal2015} employed
Str\"omgren  metallicity-sensitive indices of stars 
located in the field of the peculiar Milky Way globular cluster NGC\,2419 and found
an Fe-abundance normal spread. Nevertheless, it should be noticed that with only the 
Str\"omgren $m_{\rm 1}$  index is not possible to distinguish between the presence of an iron 
and a light-element spread, or only a light-element spread.

The paper is organized as follows: Section 2 describes the data set employed and thoroughly
establishes its accuracy. In Section 3 we derive individual [Fe/H] values for cluster stars
paying particular attention to the estimation of the metallicity uncertainties, while
Section 4 deals with the cluster metallicity distributions. Finally, Section 5 summarizes the main
conclusions of this work.

\section{Observational data}

The data used here were downloaded from the National Optical Astronomy Observatory (NOAO) 
Science Data Management (SDM) Archives\footnote{http //www.noao.edu/sdm/archives.php.}, and
are part of an observational campaign aiming at studying the age-metallicity relationships of 
Magellanic Clouds' clusters (programme ID: SO2008B-0917, PI: Pietrzynski). They consist of 
Str\"omgren $vby$ images of excellent quality (typical FWHM $\sim$ 0.6$\arcsec$) obtained on the
night of December 17, 2008, complemented with calibration and standard field images. They were 
taken with the SOAR Optical Imager (SOI) mounted on the 4.1m Southern Astrophysical Research 
(SOAR) telescope (FOV= 5.25$\arcmin$$\times$5.25$\arcmin$, scale=0.154$\arcsec$/px).
We processed all the data set according to the recipes of the SOI 
pipeline\footnote{http://www.ctio.noao.edu/soar/content/soar-optical-imager-soi}.

To standardize our photometry we measured instrumental $vby$ magnitudes of the standard
stars TYC\,7583-1011-1, 8104-856-1, 7548-698-1, 8104-820-1, HD\,3417, 58489, 66020, and 57568
\citep{hm1998,p2005}. Both Str\"omgren standard star compilations list for each star the Johnson $V$
magnitude along with the $b-y$, $m_{\rm 1}$, $c_{\rm 1}$ and $\beta$ Str\"omgren indices.  Here we 
used the $V$ magnitude and the $b-y$ and $m_{\rm 1}$ color indices to fit
the instrumental $vby$ magnitudes, so that we obtained the standard ones by simply inverting
the fitted relationships  using the {\sc iraf.photcal.inverfit}  task. 
Note that $m_{\rm 1} = (v - b) - (b - y)$. 
The stars were observed at airmass between 1.02 and 2.23
along the whole night.  We then performed fits of the expressions:\\

$v = v_1 + V_{\rm std} + v_2\times X_v + v_3\times (b-y)_{\rm std} + v_4\times m_{\rm 1 std}$,\\

$b = b_1 + V_{\rm std} + b_2\times X_b + b_3\times (b-y)_{\rm std}$,\\

$y = y_1 + V_{\rm std}  + y_2\times X_y + y_3\times (b-y)_{\rm std}$,\\

\noindent using the {\sc iraf.photcal.fitparams} routine where  $v_i$, $b_i$ and $y_i$ are the i-th fitted 
coefficients, and $X$ represents the effective airmass. Table~\ref{tab:table1} shows
the resulting coefficients. 

\begin{deluxetable*}{ccccccc}
\tablecaption{Str\"omgren transformation coefficients.\label{tab:table1}}
\tablehead{\colhead{Filter} & \colhead{coef$_{\rm 1}$} & \colhead{coef$_{\rm 2}$} & \colhead{coef$_{\rm 3}$} & \colhead{coef$_{\rm 4}$} & \colhead{rms}}
\startdata
$v$    &  1.137$\pm$0.027& 0.301$\pm$0.016& 2.008$\pm$0.058& 1.028$\pm$0.068 &  0.017 \\
       &          &      &        &       &        \\
$b$    & 0.959$\pm$0.003& 0.163$\pm$0.002 &0.942$\pm$0.003  &       &0.002        \\ 
       &          &         &        &       &        \\
$y$    & 0.946$\pm$0.015    &0.118$\pm$0.009 &-0.008$\pm$0.015  &       &0.010        \\ \hline
\enddata
\end{deluxetable*}

We obtained instrumental $vby$ magnitudes of stars located in the four cluster fields using
the stand-alone versions of  {\sc daophot}, {\sc allstar}, {\sc daomatch} and
{\sc daomaster} routine packages \citep{setal90}. The magnitudes were derived from
point-spread-function (PSF) fits performed using previously generated spatially
quadratically varying PSFs. These PSFs were modeled  from a sample of nearly 100 interactively 
selected stars distributed throughout the image, previously cleaned from fainter 
contaminating neighboring 
stars using a preliminary PSF built with nearly 40 relatively bright, well-isolated stars. 
Once we applied the resulting PSF to an image, we took advantage of the subtracted image to
identify new fainter sources which were added to the final photometric catalog. In each of the
three iterations performed, we did the PSF photometry for the whole sample of identified sources.
Finally, we transformed the instrumental magnitudes into the Str\"omgren photometric system
using the coefficients listed in Table~\ref{tab:table1}. 

The quality of our photometry was first examined in order to obtain robust estimates of the photometric 
errors. To do this, we performed artificial star tests by using the stand-alone {\sc addstar} program 
in the {\sc daophot} package \citep{setal90} to add synthetic stars, 
generated bearing in mind the color and magnitude distributions 
of the stars in the color-magnitude diagram (CMD)  - we were particularly interested in stars distributed 
along the cluster red giant branch (RGB) -, as well as 
the cluster radial stellar 
density profile. We added a number
of stars equivalent to $\sim$ 5$\%$ of the measured stars in order to avoid in the synthetic images 
significantly 
more crowding than in the original images. On the other hand, to avoid small number statistics in the
 artificial-star 
analysis, 
we created a thousand different images for each original one. We used the option of entering the number of
 photons
per ADU in order to properly add the Poisson noise to the star images. 

\begin{figure*}
     \includegraphics[width=\textwidth]{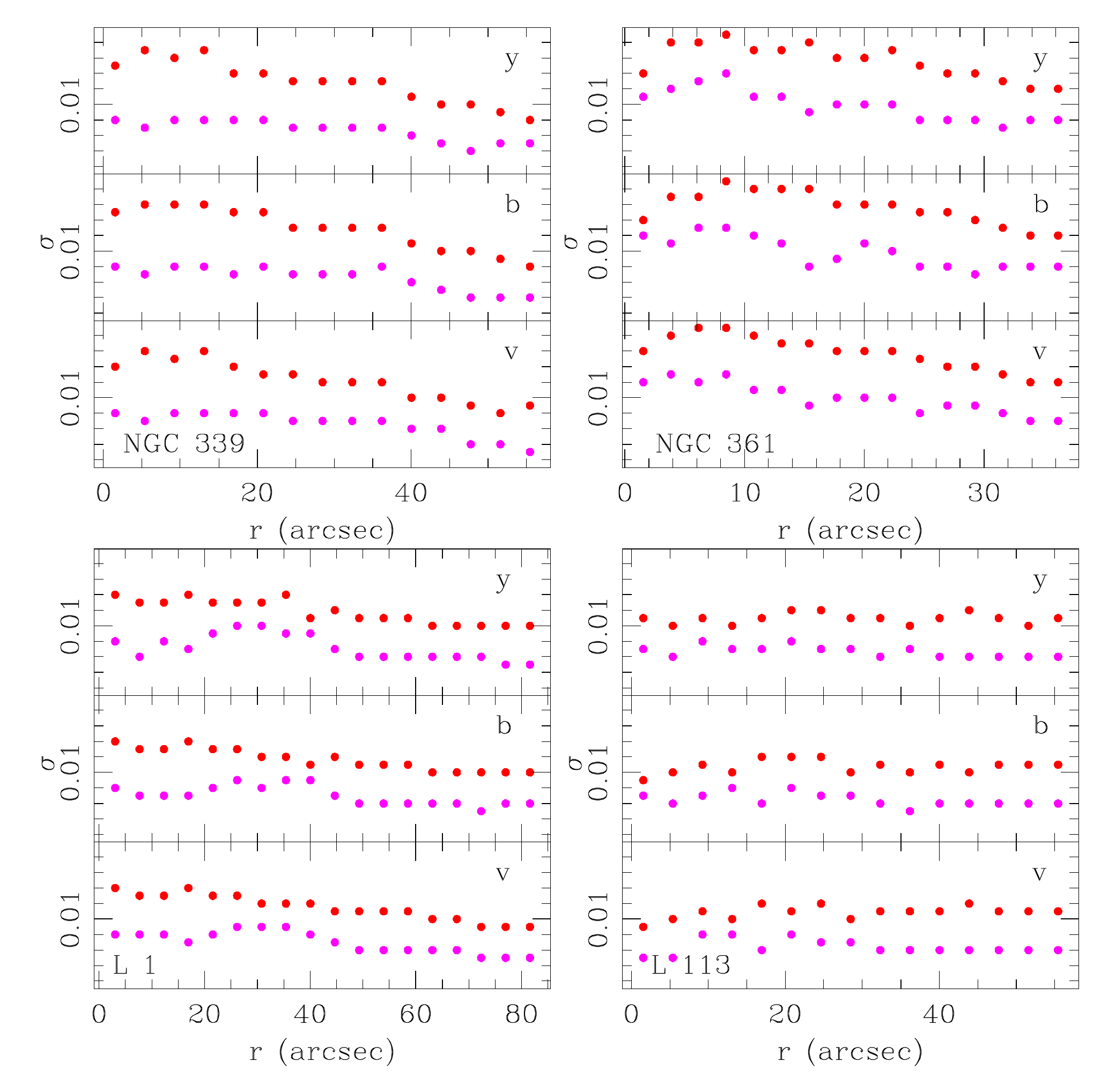}
   \caption{Photometric error estimates for the  $vby$ filters as a function of the distance
to the cluster center. Magenta and red filled circles are for $V$ = 16.5 and 18.5 mag, respectively. }
 \label{fig1}
\end{figure*}

We then repeated the same steps to obtain the photometry of the synthetic images as described above, 
i.e., 
performing three passes with the {\sc daophot/allstar} routines. 
The photometric errors were derived from the magnitude difference between the output and input data
of the added synthetic stars using the {\sc daomatch} and {\sc daomaster} tasks. We found that this difference
resulted typically equal to zero and in all the cases smaller than 0.003 mag. The respective rms errors were 
adopted as the photometric errors.
Fig.~\ref{fig1} illustrates the behavior of these errors as a function of the distance from the
cluster center for two different magnitude level, namely $V$ = 16.5 mag and 18.5 mag, respectively. 
These magnitudes
roughly correspond to the upper and lower limits of the cluster RGBs used in this work
(see Figs.~\ref{fig2}-\ref{fig5}).

\section{Metallicity estimates}

As for metalliciy estimates of individual stars we used the semi-empirical 
calibration of the reddening free
metallicity-sensitive index $m_{\rm 1}$$_{o}$, in turn based on the stars' $(v-y)_o$ colors, 
derived by \citet{calamidaetal2007}. Particularly, we used the expression:

\begin{equation}
{m_{\rm 1}}_o = \alpha + \beta {\rm [Fe/H]} + \gamma (v-y)_o + \delta {\rm [Fe/H]}(v-y)_o
\end{equation}

\noindent where $\alpha$ = -0.309, $\beta$=-0.090$\pm$0.002, $\gamma$=0.521$\pm$0.001, and
$\delta$=0.159$\pm$0.001, respectively. We adopted the semi-empirical relation, since this is 
recommended by \citet{calamidaetal2007} as being the most robust for estimating the metallicity 
of red giants (see also \citet{calamidaetal2009,adenetal2009,arnaetal2010,franketal2015}). 
Indeed, \citet{pk2018}
have recently obtained Str\"omgren-based [Fe/H] values from 10 LMC ancient Large Magellanic Cloud 
globular clusters over a broad metallicity range (-2.1 $\le$ [Fe/H] (dex) $\le -1.0 $) 
in excellent agreement with mostly high-dispersion spectroscopic values.  Note that our derived
mean cluster metallicities (see Table~\ref{tab:table2}) are also in excellent agreement with spectroscopic
values). We provide in the Appendix
a comparison of the [Fe/H] values derived for selected stars (see below) from the semi-empirical calibration with those based on the empirical and theoretical ones, respectively. The reddening corrected ${m_{\rm 1}}_o$ and $(v-y)_o)$ colors were obtained
from the $E(m_{\rm 1})$ and $E(v-y)$ color 
excesses computed using the $E(X)/E(B-V)$ ratios given by \citet{cm1976} and the $E(B-V)$ color excesses
obtained from the NASA/IPAC  Extragalactic Data 
base\footnote{http://ned.ipac.caltech.edu/. NED is operated by the Jet Propulsion Laboratory, California 
Institute of Technology, under contract with NASA.} (NED) (see Table~\ref{tab:table2}).

To estimate the uncertainties in the [Fe/H] values,
we performed a full analytical propagation of errors, including those on the
calibration coefficients, as follows:\\

$\sigma({\rm [Fe/H]})^2 = \left(\frac{\partial {\rm [Fe/H]}}{\partial \alpha} \sigma(\alpha)\right)^2 
+ \left(\frac{\partial {\rm [Fe/H]}}{\partial \beta} \sigma(\beta)\right)^2 +
\left(\frac{\partial {\rm [Fe/H]}}{\partial \gamma}\sigma(\gamma)\right)^2 + 
  \left(\frac{\partial {\rm [Fe/H]}}{\partial \delta}\sigma(\delta)\right)^2 + 
  \left(\frac{\partial {\rm [Fe/H]}}{\partial {m_{\rm 1}}_o}\sigma({m_{\rm 1}}_o)\right)^2 + 
\left(\frac{\partial {\rm [Fe/H]}}{\partial (v-y)_o}\sigma((v-y)_o)\right)^2$,\\

\vspace{0.5cm}

$\sigma({\rm [Fe/H]})^2 = \left(\frac{0.002{\rm [Fe/H]}}{c}\right)^2 + 
\left(\frac{0.001(v-y)_o}{c}\right)^2 + \left(\frac{0.001{\rm [Fe/H]}(v-y)_o}{c}\right)^2 +
\left(\frac{\sigma({m_{\rm 1}}_o)}{c}\right)^2 + \\
\left(\frac{(-0.521c - 0.159({m_{\rm 1}}_o +0.309 -0.521(v-y)_o)\sigma((v-y)_o)}{c^2}\right)^2$

\vspace{0.5cm}

\noindent where c = $-0.090 + 0.159(v-y)_o$, and $\sigma({m_{\rm 1}}_o)$ and $\sigma((v-y)_o)$
are the photometric errors in ${m_{\rm 1}}_o$ and $(v-y)_o$, respectively, according to
the position of the stars with respect to the cluster's center (see Fig.~\ref{fig1}).

We then carefully selected {\it bonafide} cluster members on the basis of the following criteria: 
i) they are located inside the clusters' radii \citep{hz2006,getal09,getal11}. ii) They are 
distributed along the RGBs above the respective red clumps/horizontal branches.
The cluster RGBs in the $V$ versus $b-y$ and $V$ versus $v-y$ CMDs are not
dominated by metallicity effects, so that they result in narrow star sequences.
We traced the RGB ridge lines and discarded any stars with $b-y$ and $v-y$ colors that
differ in more than 0.05 mag from those of the RGC ridge lines.
In addition, the cluster RGBs are the least contaminated CMD features by field stars. Indeed,
by using different field regions of equal cluster areas, distributed around the clusters, we
found that the number of field stars  that fall inside the RGB strips is smaller
than 10 per cent with respect to the total number of adopted RGB cluster members. 
iii) They span the readily visible cluster [Fe/H] range as judged by the
dispersion ($\sim$ within 2$\times$standard deviation) of [Fe/H] values for the brightest 
selected stars, i.e., those drawn with magenta open circles in  Figs.~\ref{fig2}-\ref{fig5}.
 In order to achieve this, we 
discarded some few stars that only satisfied criteria i and ii, but fall outside the
aforementioned [Fe/H] range. They are drawn with filled
red circles in Figs.~\ref{fig2}-\ref{fig5}. Note that the brightest selected stars
(magenta circles) are those with more accurate photometry, and hence with the
smallest [Fe/H] uncertainties. Likewise, as the $V$ magnitude increases, both the metallicty 
errors and the dispersion of the individual [Fe/H] values increase (see bottom-right
panels of Figs.~\ref{fig2} to \ref{fig5}), because of the poorer photometry quality
(see Fig.~\ref{fig1}). For these
reasons, the observed [Fe/H] range of the brightest selected stars more properly reveals the
cluster metallicity range. Because of this metallicity range should be the same at any magnitude 
level, we used it for discarding some few stars fainter than those
marked with magenta circles. We simply included the stars with red circles with the purpose of
enlarging the star sample, while the clusters' mean metallicities and dispersions keep
unchanged from an astrophysical point of view with respect the values derived using only the brightest selected stars. 

Figs.~\ref{fig2} to \ref{fig5} show the
resulting observed CMDs and the derived individual [Fe/H] values with their respective uncertainties.
Stars observed within the cluster areas and those for comparison star fields with equal
cluster areas are depicted with filled black and green circles, while he selected 
stars are encircled
with magenta and red open circles, respectively. As can be seen, the combination of 
an accurate photometry (see also Table~\ref{tab:table1} and Fig.~\ref{fig1}) and a
accurate metallicity calibration (see eq.(1)) resulted in an advantageous tool for
estimating cluster RGB stars' metallicities with uncertainties that, in the case of
the brightest objects, are of the same order than those expected from high-dispersion
spectroscopy. For the sake of the reader we have also included the $v-y$ versus $m_{\rm 1}$
color-color diagrams in Fig.~\ref{fig6}.

\section{Analysis and discussion}

In order to assess whether the metallicty distributions are shaped by the presence of
an intrinsic spread in Fe-abundances,
we derived the mean and dispersion of each cluster's Fe-abundance by
employing a maximum likelihood approach.
The relevance lies in accounting for individual star measurements, which could 
artificially inflate the dispersion if ignored.
We optimized the probability $\mathcal{L}$ that a
given ensemble of stars with 
metallicities [Fe/H]$_i$ and errors $\sigma_i$ are drawn from a population with mean
Fe-abundance $<$$[Fe/H]$$>$ and dispersion  W
\citep[e.g.,][]{pm1993,walkeretal2006}, 
as follows:\\

\begin{figure}
   \includegraphics[width=\columnwidth]{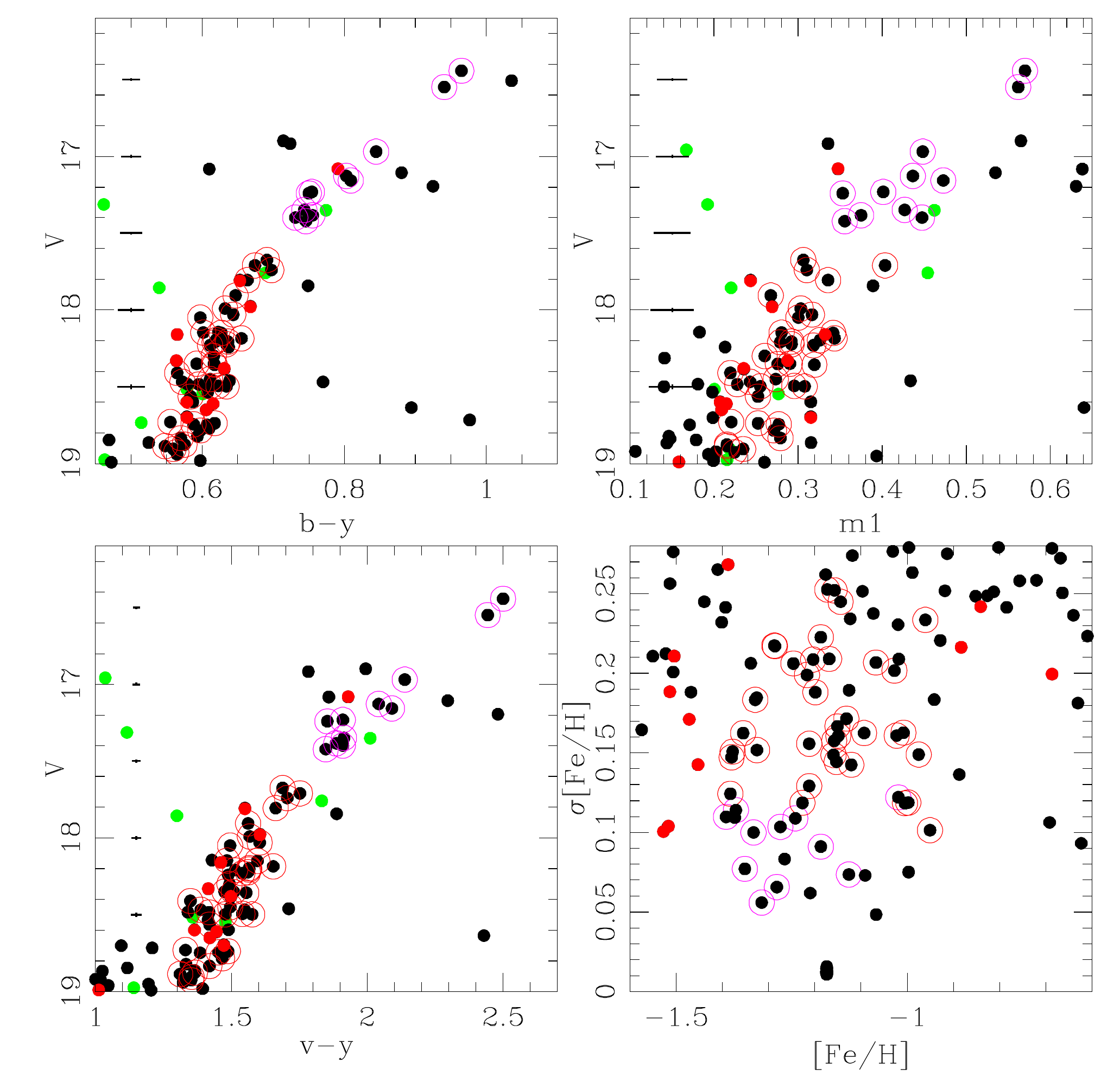}
    \caption{Observed CMDs of stars observed in the field of NGC\,339. Black and
green filled circles represent stars observed in the cluster area and in
a comparison star field with an equal cluster area, respectively. Red filled circles 
represent stars that only satisfied the selection criteria i and ii (see text for details).
Average error bars are also depicted at the left-hand margin of the CMDs.
 The individual metallicities 
([Fe/H]) and their respective uncertainties are depicted in the bottom-right panel. We split the
sample in brighter and fainter selected stars highlighted with magenta and red open circles,
respectively.}
   \label{fig2}
\end{figure}

\begin{figure}
   \includegraphics[width=\columnwidth]{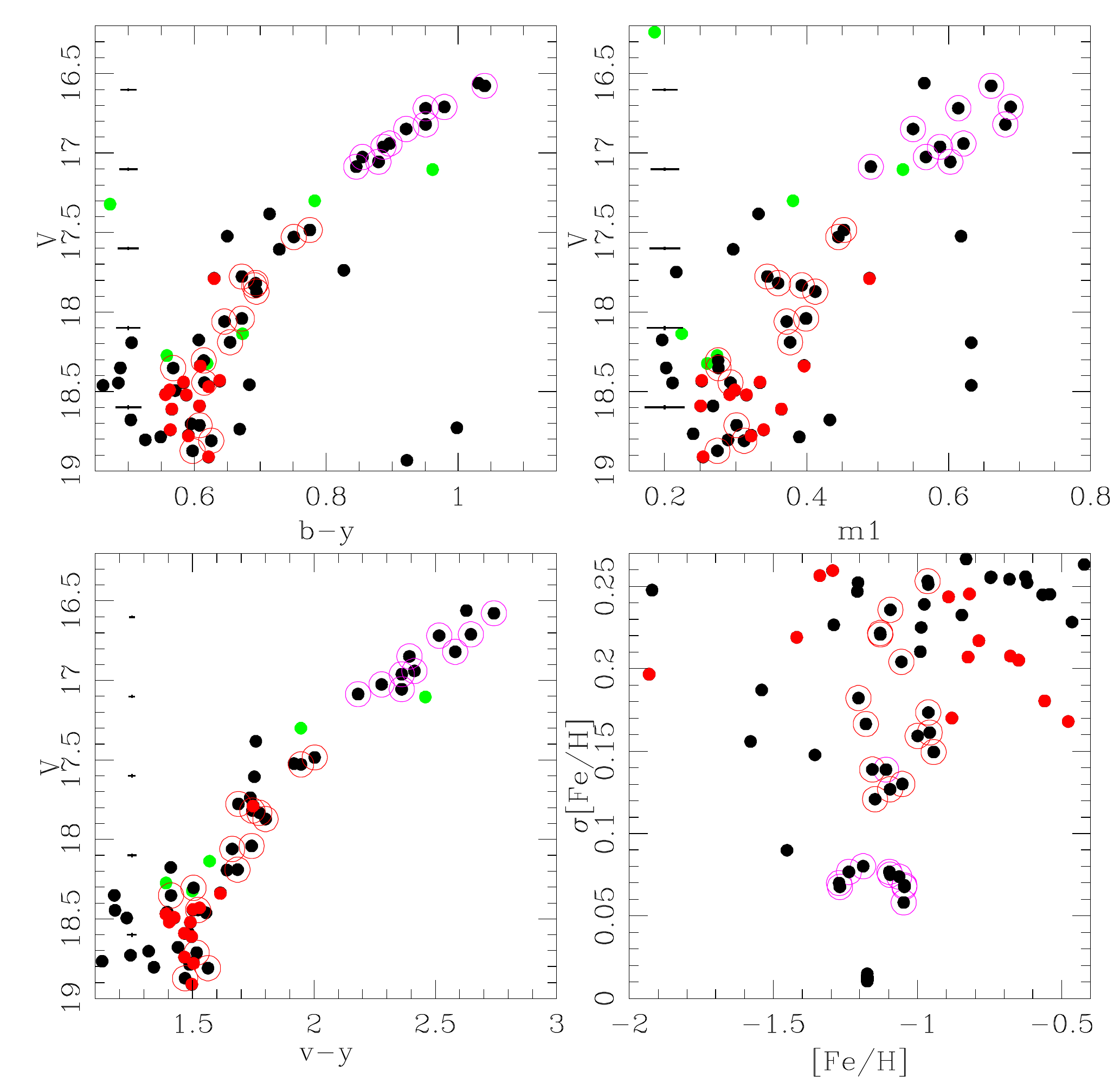}
    \caption{Same as Fig.~\ref{fig1} for NGC\,361.}
   \label{fig3}
\end{figure}

\begin{figure}
   \includegraphics[width=\columnwidth]{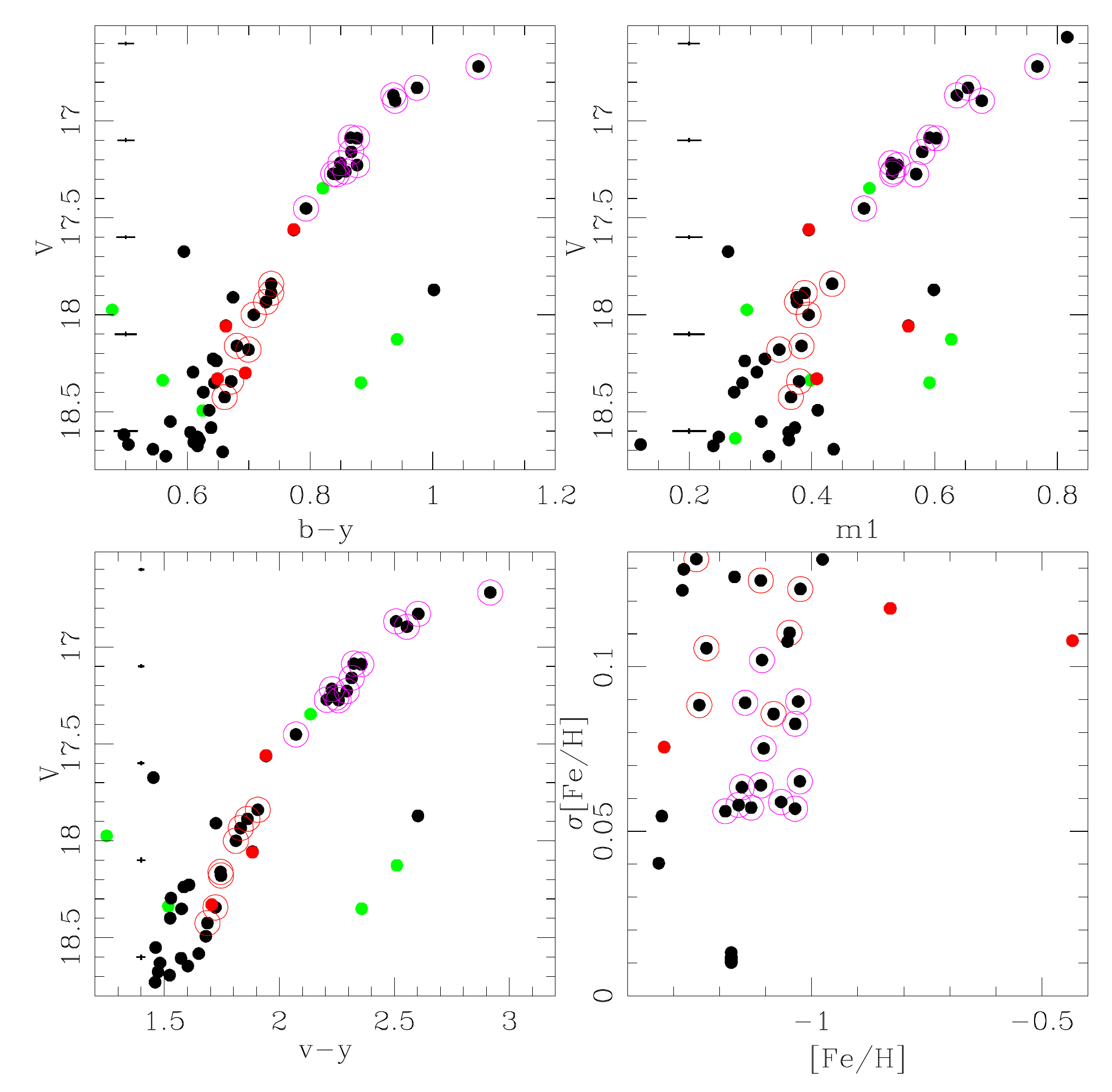}
    \caption{Same as Fig.~\ref{fig1} for Lindsay\,1.}
   \label{fig4}
\end{figure}

\begin{figure}
   \includegraphics[width=\columnwidth]{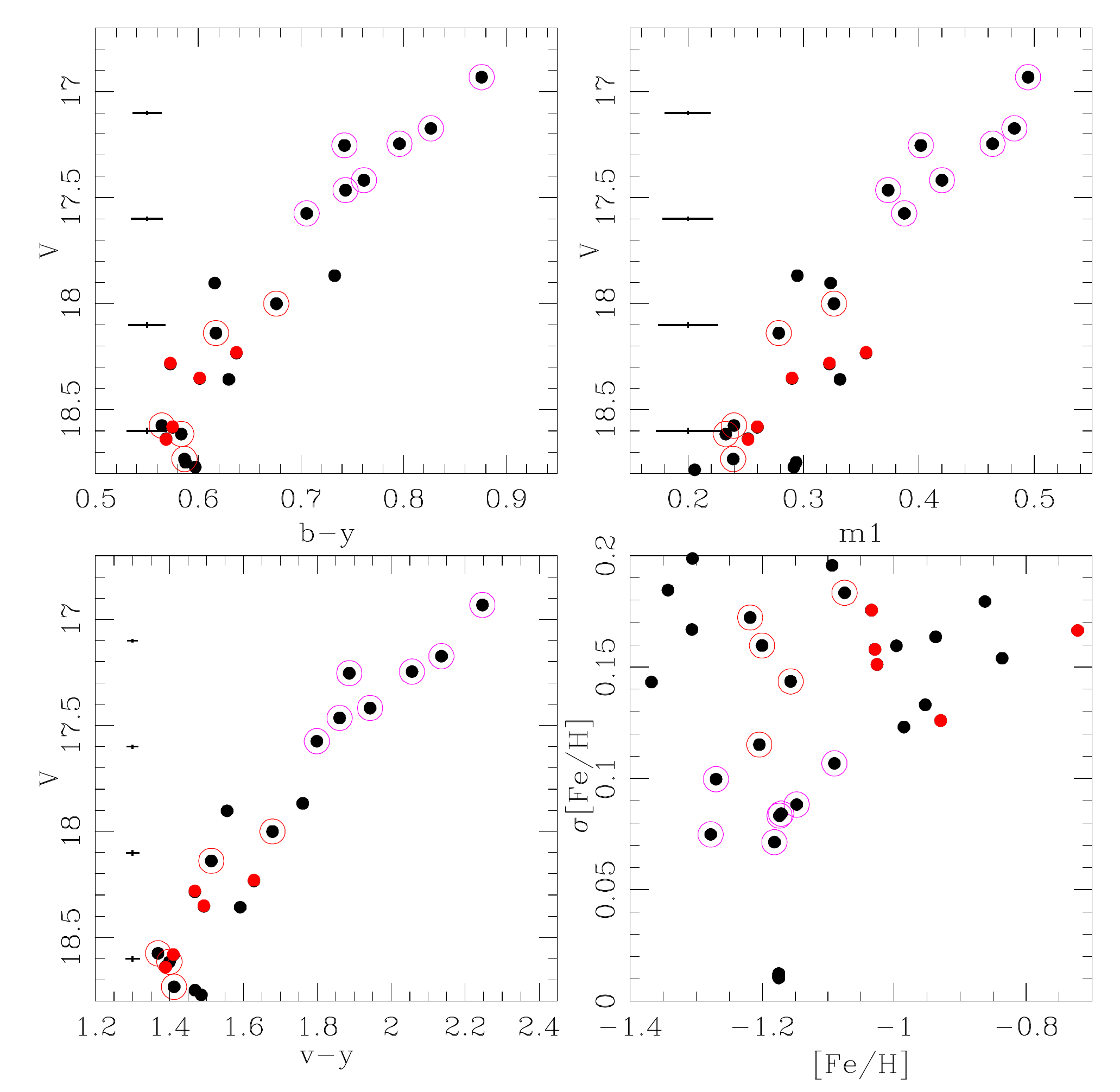}
    \caption{Same as Fig.~\ref{fig1} for Lindsay\,113.}
   \label{fig5}
\end{figure}

\begin{figure}
   \includegraphics[width=\columnwidth]{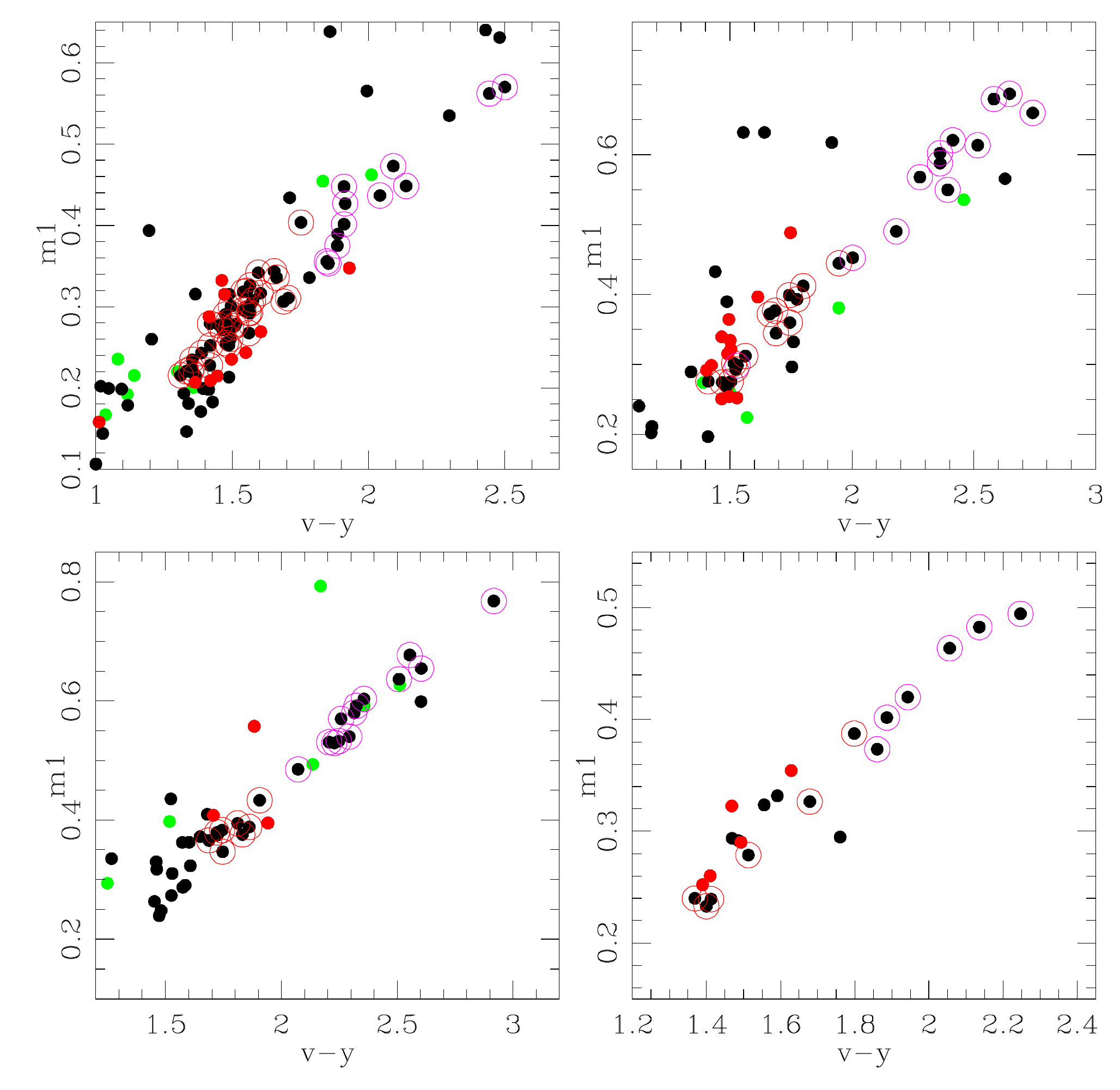}
    \caption{$v-y$ versus $m_{\rm 1}$ color-color diagrams of NGC\,339 (upper-left),
NGC\,361 (upper-right), Lindsay\,1 (bottom-left), and Lindsay\,113 (bottom-right),
respectively. Symbols as are in Fig.~\ref{fig1}.}
   \label{fig6}

\end{figure}

$\small
\mathcal{L}\,=\,\prod_{i=1}^N\,\left( \, 2\pi\,(\sigma_i^2 + W^2 \, ) 
\right)^{-\frac{1}{2}}\,\exp \left(-\frac{([Fe/H]_i \,- <[Fe/H]>)^2}{2(\sigma_i^2 + W^2)}
\right) \\$

\noindent where the errors on the mean and dispersion were computed from the respective 
covariance matrices. 
Table~\ref{tab:table2} lists in the last columns all our results, where the W values refer 
to the selected N brightest stars (open magenta circles in Figs.~\ref{fig2}-\ref{fig5}).
In case of using all the cluster RGB stars, W turns out to be equal to zero for all the
clusters. This is a consequence of dealing with larger [Fe/H] errors, since both
groups of selected stars (brighter and fainter than a fixed $V$ magnitude) span nearly the
same metallicity range and have similar average metallicity values. Here,
the larger individual [Fe/H] errors of fainter stars blur any possible intrinsic spread.

Our findings show that among the two studied MP SMC clusters only NGC\,339 exhibits
a relatively small Fe-abundance dispersion. This result could imply that the
intrinsic spread in metallicity is not a common feature in these MP SMC star clusters, as is
not in Milky Way old globular clusters either, or that more accurate
metal abundance estimates are required. In any case, the amount of metallicity
spread would appear to be somehow marginal, i.e., smaller than the values found in eight
Milky Way globular clusters ($>$ 0.10 dex). Likewise, note also that we have not used 
some RGB stars that are located inside the clusters' radii, are placed along the cluster 
RGBs and have similar [Fe/H] errors like those selected stars, which would 
enlarge the cluster metallicity range. In this sense, the Fe-abundance spreads found in 
this study should be consider as lower limits. 

The Str\"omgren metallicities derived in this analysis come from measurements of the
$m_{\rm 1}$ index = ($v-b$) - ($b-y$) calibrated in terms of [Fe/H] values obtained from
high-dispersion spectroscopy. In this sense, \citet{calamidaetal2007} used the $m_{\rm 1}$ 
index as a photometric proxy for the iron abundance. However, the CN absorption band at
4142\,\AA\, is near to the effective wavelength of the $v$ filter, so that
the derived dispersions W could reflect variations in light-element abundances.
Nevertheless, according to \citep{limetal2017}, globular clusters with 
heavy element abundance variations do show a CN-CH positive correlation, unlike normal
globular clusters. In this sense, the derived [Fe/H] values might reflect both heavy and light
element abundances or only those from light elements. Further analyses with the Str\"omgren
$c_{\rm 1}$ index or from spectroscopic data are needed.
 
For NGC\,361 we obtained an W value similar to that for the MP SMC cluster NGC\,339, from
which we infer that the former could also be added to the list of SMC clusters
harboring MPs. In the case of Lindsay\,113, we cannot conclude on the existence of
MPs, because of the lack of evidence of any small mean iron spread among the
measured stars. Several theoretical models have recently proposed different scenarios
to describe abundance anomalies in a variety of chemical elements in massive
clusters with MPs
\citep[see, e.g.][and references therein]{bt2016,bekki2017,bailin2018,gielesetal2018,kimeal2018}.
They have been mainly stimulated by observational results of
anti-correlations between chemical abundances of light elements, although
some of the models also suggest mechanisms to obtain
intrinsic [Fe/H] spreads $>$ 0.05 dex. Here we would like to mention that our studied 
clusters are more massive than the MP globular cluster ESO\,452-SC11 
($M=$6.8$\pm$3.4 $\times$10$^3M_\odot$)
(see Table~\ref{tab:table2}, for which \citet{simpsonetal2017} found difference in
chemical compositions.
 
\begin{deluxetable*}{lcccccccccc}
\tablecaption{Astrophysical properties of SMC clusters.\label{tab:table2}}
\tablehead{\colhead{Cluster} & \colhead{$E(B-V)_{\rm NED}$} & \colhead{Age} & \colhead{Ref.} & 
\colhead{[Fe/H]$_{\rm lit.}$} & \colhead{Ref.} & \colhead{Mass} & \colhead{Ref.} & \colhead{[Fe/H]} & \colhead{W} &
\colhead{N}\\
   \colhead{ }  &     \colhead{(mag)}  & \colhead{(Gyr)} &\colhead{ } & \colhead{(dex)} & \colhead{ } &
 \colhead{($\times$10$^5M_\odot$)} & \colhead{} & \colhead{}}
\startdata
NGC\,339     & 0.03 & 5.4$\pm$1.0 & 1,2   & -1.19$\pm$0.10 & 4,5   & 0.8 &  8 &  -1.27$\pm$0.03 & 0.04$\pm$0.03 & 11\\
NGC\,361     & 0.03 & 6.8$\pm$0.5 & 6     & -1.08$\pm$0.10 & 5     & 2.0 &  8 & -1.13$\pm$0.03 & 0.06$\pm$0.03 & 10\\
Lindsay\,1   & 0.03 & 7.5$\pm$0.5 & 1,6   & -1.01$\pm$0.10 & 5     & 2.0 &  7 & -1.11$\pm$0.02 & 0.00$\pm$0.04 & 13\\
Lindsay\,113 & 0.04 & 4.6$\pm$1.0 & 3,6   & -1.17$\pm$0.10 & 4     & --- &    & -1.19$\pm$0.03 & 0.00$\pm$0.04 & 7\\
\enddata
\tablecomments{ Ref.: (1) \citet{glattetal2008a}; (2) \citet{p11c}; (3) \citet{petal07}; (4) \citet{dh98}; 
(5) \citet{mucciarellietal2009a}; (6) \citet{metal98}; (7) \citet{getal11}; (8) \citet{mvdm05}.}
\end{deluxetable*}

\section{Conclusion}

With the aim of investigating whether SMC intermediate-age clusters with
light chemical abundance variations also shows Fe-abundance spreads,
we conducted an analysis of NGC\,339 and Lindsay\,1, recently
confirmed as MP SMC clusters. We also added NGC\,361 and Lindsay\,113,
two SMC intermediate-age clusters with well-populated RGBs.

We made use of publicly available $vby$ Str\"omgren images centered on these
clusters, from which we obtained accurate photometric data sets. We
extensively and carefully probed the precision of our photometry, focused
on stars distributed along the cluster RGBs, for which we attained 
uncertainties smaller than 0.02 magnitudes in all the three filters,
regardless the stars are located in the innermost or outermost regions
of the clusters.

As the metallicity estimate is concerned, we employed a well-established,
high-dispersion spectroscopy based calibration of the Str\"omgren
metallicity-sensitive index $m_{\rm 1}$.  We derived individual [Fe/H] values 
for RGB stars brigher that those placed at the cluster red clump/horizontal branch.
For the brightest cluster RGC stars we estimated uncertainties comparable to those
expected from precise spectroscopic observations. From these metallicity
estimates we computed their meam and intrinsic dispersion. We obtained
mean metal abundances in very good agreement with those previously 
derived from spectroscopic studies and small intrinsic Fe-abundance
spreads in NGC\,339 and NGC\,361. From this results we speculated with
the possibility of NGC\,361 being the fifth SMC intermediate-age cluster
with MPs. The lack of detection of significant metallicity spreads in
the MP SMC cluster Lindsay\,1 as well as in Lindsay\,113 could be
attributed to the fact that this is not a common phenomenon among
MP star clusters, or we need  even more accurate metallicity
estimates in order to asses the cluster metallicity spread more reliably.

\acknowledgments

Based on observations obtained at the Southern Astrophysical Research (SOAR) telescope,
 which is a joint project of the Minist\'{e}rio da Ci\^{e}ncia, Tecnologia, 
Inova\c{c}\~{o}es e Comunica\c{c}\~{o}es (MCTIC) do Brasil, the U.S. National
 Optical Astronomy Observatory (NOAO), the University of North Carolina at Chapel Hill (UNC),
 and Michigan State University (MSU).
We thank the referee for the thorough reading of the manuscript and
timely suggestions to improve it.

\bibliographystyle{aasjournal}

\appendix

\section{[Fe/H] values based on Calamida et al (2007)'s calibrations.}

\begin{figure}
   \includegraphics[width=\columnwidth]{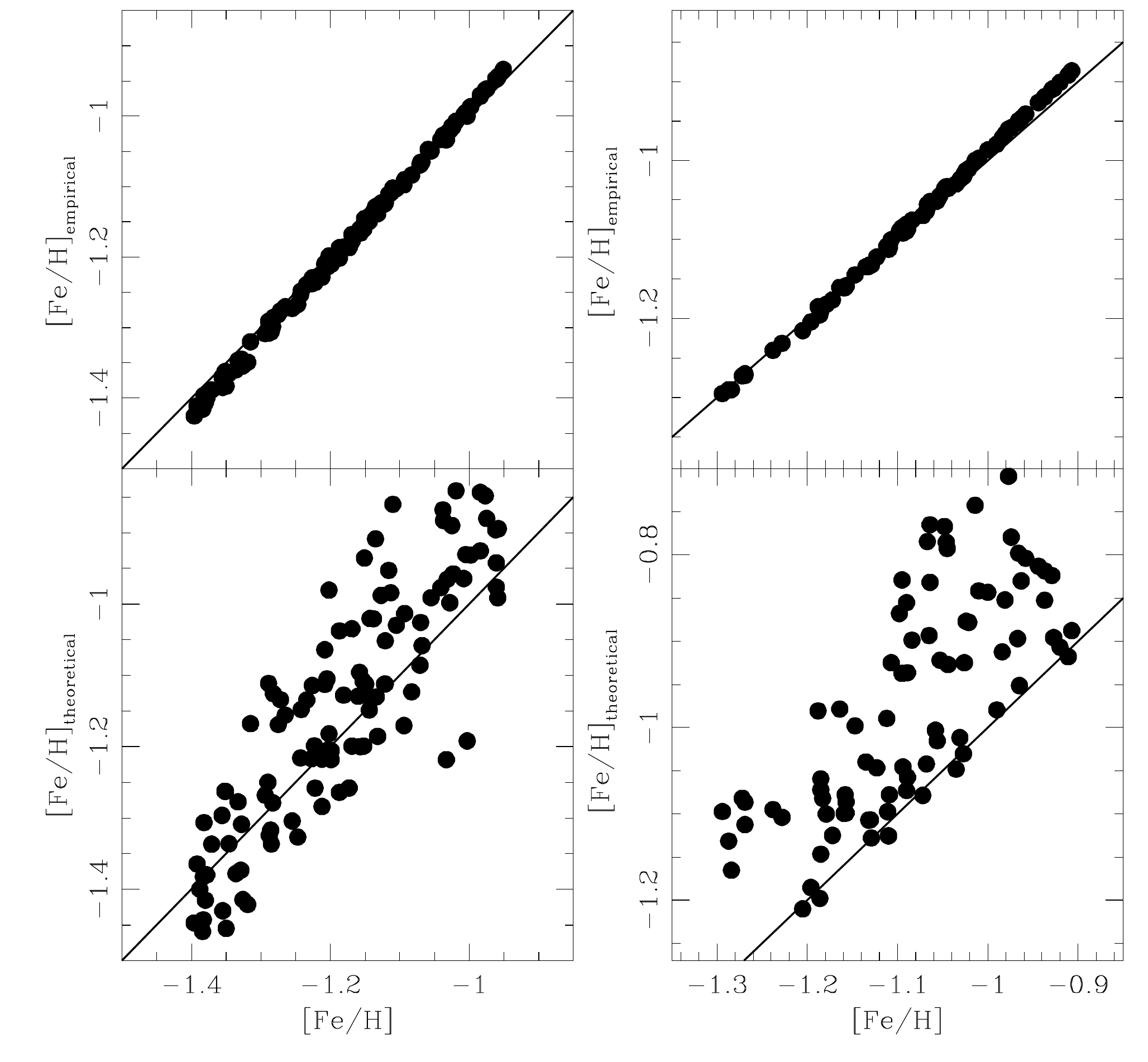}
    \caption{[Fe/H] values derived from the semi-empirical calibration of \citet{calamidaetal2007} compared with those based on the empirical (top) and theoretical (bottom) calibrations, for 
NGC\,339 (left) and NGC\,361 (right). The solid line represents the identity relationship.}
   \label{fig1A}
\end{figure}

\begin{figure}
   \includegraphics[width=\columnwidth]{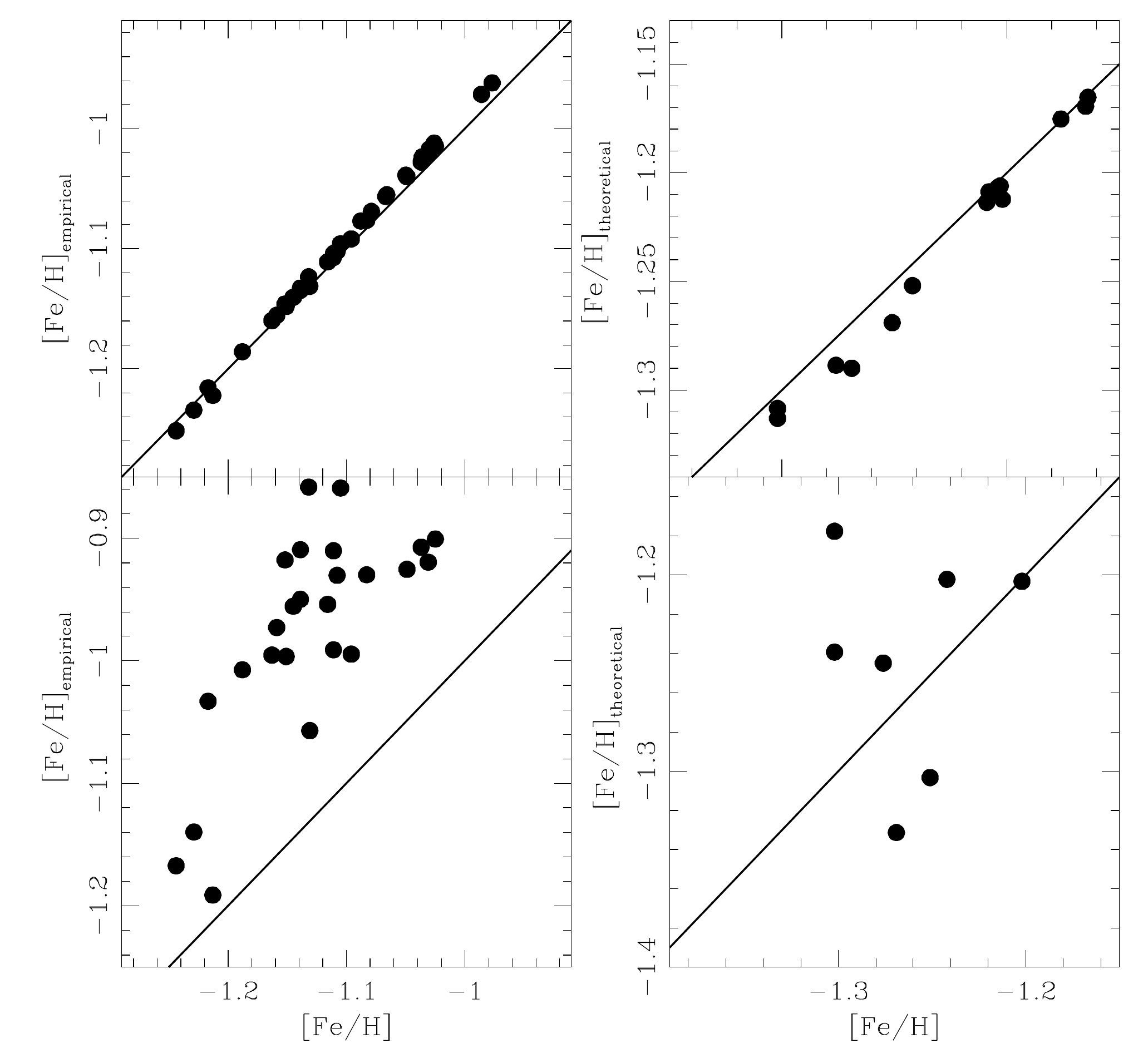}
    \caption{[Fe/H] values derived from the semi-empirical calibration of \citet{calamidaetal2007} compared with those based on the empirical (top) and theoretical (bottom) calibrations, for 
L\,1 (left) and L\,113 (right).The solid line represents the identity relationship.}
   \label{fig1B}
\end{figure}

For the sake of the reader, we present here a comparison between metallicities derived from 
different \citet{calamidaetal2007}'s calibrations, namely: empirical, theoretical and semi-empirical ones. 
As can be seen in Figs.~\ref{fig1A} and \ref{fig1B}, metallicities derived with the empirical or semi-empirical
calibrations agree quite well, and the present results 
would not change if a different calibration were used to estimate the
star metal abundances. The theoretical calibration underestimates
metallicities by $\sim$ 0.2 dex, in particular in the intermediate
metallicity range of the considered clusters, as shown by \citet{calamidaetal2007}.

\end{document}